\documentclass[%
 reprint,
superscriptaddress,
 amsmath,amssymb,
 aps,
 prd,
]{revtex4-1}

\usepackage{fancyhdr}
\usepackage{graphicx}
\usepackage{tikz-cd} 
\usepackage{tcolorbox}
\usepackage{tensor}
\usepackage{tikz,relsize,booktabs}

\usepackage{xcolor}
\definecolor{tensorblue}{rgb}{0.8,0.8,1}
\definecolor{tensorred}{rgb}{1,0.5,0.5}
\definecolor{tensorpurp}{rgb}{1,0.5,1}

\usepackage{tikz}

\tikzset{tens/.style={fill=tensorblue}}
\tikzset{diag/.style={fill=green!50!black!50}}
\tikzset{isom/.style={fill=orange!30}}
\tikzset{proj/.style={fill=tensorred}}

\tikzset{tengrey/.style={fill=black!20}}
\tikzset{tenpurp/.style={fill=tensorpurp}}

\usetikzlibrary{calc}

\usepackage{amsfonts}

\newcommand{\Tr}{{\rm Tr\;}}

\def\bec{\begin{center}}
\def\eec{\end{center}}
\def\beq{\begin{equation}}
\def\eeq{\end{equation}}
\def\bea{\begin{eqnarray}}
\def\eea{\end{eqnarray}}

\begin{document}
\title{Tensor network formulation of the massless Schwinger model}

\author{Nouman Butt}
\email{ntbutt@syr.edu}
\affiliation{Department of Physics, Syracuse University, Syracuse NY 13244}
\author{Simon Catterall}
\affiliation{Department of Physics, Syracuse University, Syracuse NY 13244}
\author{Yannick Meurice}
\affiliation{Department of Physics and Astronomy, The University of Iowa, Iowa City, IA 52242, USA}
\author{Judah Unmuth-Yockey}
\email{jfunmuthyockey@gmail.com}
\affiliation{Department of Physics, Syracuse University, Syracuse NY 13244}

\begin{abstract}
We construct a tensor network representation of the partition function
for the massless Schwinger model on a two dimensional lattice using staggered fermions. The tensor network representation
allows us to include a topological term. Using a particular
implementation of the tensor renormalization group (HOTRG) we calculate the phase diagram of the
theory. For a range of values of the coupling to the topological term $\theta$ and the gauge coupling $\beta$ we compare with results from
hybrid Monte Carlo when possible and find good agreement.
\end{abstract}
\maketitle
\section{Introduction}

In recent years there has been a surge of interest
in applying tensor network methods to calculate the properties of lattice spin and gauge models \cite{PhysRevD.88.056005,Kadoh2019,PhysRevD.99.114507,PhysRevD.99.074502,PhysRevD.98.094511,qpotts,PhysRevD.100.054510,Wang_2014,PhysRevE.89.013308}. In low dimensions
these formulations can avoid the usual sign problems associated with negative or complex probability weights that plague Monte Carlo approaches, and can yield
very efficient computational algorithms \cite{PhysRevD.89.016008,PhysRevLett.115.180405,PhysRevLett.99.120601,PhysRevB.86.045139}. For compact fields the general strategy has been to employ
character expansions for all Boltzmann
factors occurring in the partition function and subsequently to integrate out the original fields, yielding an equivalent formulation in terms of integer---or half-integer---valued fields. Typically local tensors can be built from these discrete variables
and the partition function recast as the full contraction of all tensor indices. 

However, writing local tensors for models with relativistic lattice fermions is more complicated~\cite{10.1093/ptep/ptx080,PhysRevD.90.014508,10.1093/ptep/ptv022,Kadoh2018}.
One reason is
tied to the Grassmann nature of the fermions which can induce additional, non-local sign problems which may
be hard to generate from local tensor contractions. However, 
Gattringer et. al. have shown in Ref.~\cite{GATTRINGER2015732} that a suitable dual formulation can
be derived in the case of the massless Schwinger model which is free of these sign problems. Using this dual
representation they have formulated a general Monte Carlo algorithm that can be used to simulate
the model even in the presence of non-zero chemical potential and topological terms \cite{GOSCHL201763}.

Other directions into the investigation of the Schwinger model have appeared in recent years as well.  One approach has been the use of other numerical renormalization group methods like the density matrix renormalization group (DMRG) with matrix product states or matrix product operators (MPS or MPO).  The massive Schwinger model with staggered fermions was investigated in Ref.~\cite{PhysRevD.66.013002} using the DMRG.  In Ref~\cite{Banuls2013} the mass spectrum of the Schwinger model was calculated at zero and finite mass, and in Ref.~\cite{PhysRevD.94.085018} the authors studied the Schwinger model at finite temperature using the DMRG with MPO.  The effect of truncation on the number of representations retained in the electric field basis for the Schwinger model was investigated in Ref.~\cite{PhysRevD.95.094509}.  In Ref.~\cite{PhysRevD.100.036009} the confinement properties of the Schwinger model in the presence of a topological term were studied, and in Ref.~\cite{lena_topo} the authors considered the effects of a topological term on the vacuum structure of the model, again using the DMRG.

Similarly, Ref.~\cite{PhysRevD.98.074503} looked at a $\mathbb{Z}_{n}$ formulation of the Schwinger model using the DMRG.  They found that at large $n$, one recovers similar results to the original continuous $U(1)$ symmetry in the Schwinger model.  Out of equilibrium properties were looked at in Ref.~\cite{giuseppe_znqed} for that same model.

On top of that, proposals and investigations into the potential for quantum simulations and computations of the Schwinger model were done in Refs.~\cite{PhysRevLett.112.201601,Muschik_2017,PhysRevA.98.032331}.  In Ref.~\cite{PhysRevLett.112.201601} the lattice Schwinger model was considered for quantum simulation using cold atoms in an optical lattice.  In Ref.~\cite{Muschik_2017} the authors considered general $U(1)$ lattice gauge theories and they integrate out the gauge degrees of freedom, being left with a model of strictly matter, interacting non-locally.  This model would be implemented using trapped ions.  In Ref.~\cite{PhysRevA.98.032331}, the authors considered the joint computation of the lattice Schwinger model using classical and quantum computers.   

In this paper we show that the dual world-line formulation from Ref.~\cite{GATTRINGER2015732} can be replicated by contraction of a suitable tensor network.  It should be noted that a tensor formulation of the model allows for the definition of a transfer matrix, quantum Hamiltonian, and local Hilbert space.
Rather than following a Monte Carlo strategy we instead use and follow the philosophy of the tensor
renormalization group to coarse grain this tensor network.  From this we calculate the partition function and free energy.
We show that the results agree well with both Ref.~\cite{GOSCHL201763} and conventional hybrid Monte Carlo simulations where
the latter can be performed.

We start by reviewing the construction of the dual representation and show 
how the resulting dimer/loop representation can be obtained by the contraction of a suitable tensor network and derive the
form of the fundamental tensor that is needed. We then describe the results of a coarsening of this tensor
network using the HOTRG algorithm, calculate the free energy and its derivatives and compare the results
to Monte Carlo simulations. We then go on to add a topological term to the action with coupling $\theta$.
We conclude with
a summary of the advantages and disadvantages of the method in this context.

\section{Massless Schwinger Model and its dual representation}

We begin with the one-flavor staggered action for the massless Schwinger model on a $N_{s} \times N_{\tau}$ lattice with action
\begin{equation}
    S = S_{F} + S_{g}
\end{equation}
with
\begin{align}
\nonumber
    S_{F} = \frac{1}{2} \sum_{x, \mu} \eta_{\mu}(x) &[ \bar{\psi}(x) U_{\mu}(x) \psi(x+ \mu) \\
    &- \bar{\psi}(x+\mu)U^{\dagger}_{\mu}(x) \psi(x) ]
\end{align}
and
\begin{equation}
S_{g} = - \beta \sum_{x} {\rm Re}\, [ U_P(x) ],
 \end{equation}
where the Abelian gauge field $U_\mu(x)=e^{i A_\mu(x)}$ lives on the link between lattice sites $x$ and $x+\mu$
and the fermions $\psi(x)$ and $\bar{\psi}(x)$ live at the sites. $U_P$ is the usual Wilson plaquette operator 
 $U_P(x) =  \sum_{\mu < \nu }U_{\mu}(x) U_{\nu}(x+\mu) U^{\dagger}_{\mu} (x + \nu) U^{\dagger}_{\nu}(x) $.
 The partition function for this model is then given by 
\begin{align}
\nonumber
Z  &= \int D[U] D[\bar{\psi}]D[\psi] \; e^{-S} \\
   &=\int D[U] e^{\beta \sum_x {\rm Re} [U_P(x)]}Z_F(U)
\end{align}
with $ \int D[U] = \prod_{x} \int_{-\pi}^{\pi} dA_{\mu}(x) / 2\pi $, $ \int D[\bar{\psi}] D[\psi] = \prod_{x} \int d\bar{\psi}(x) d\psi(x) $, and $Z_{F}$ represents the part of the partition function that depends on the fermion fields.

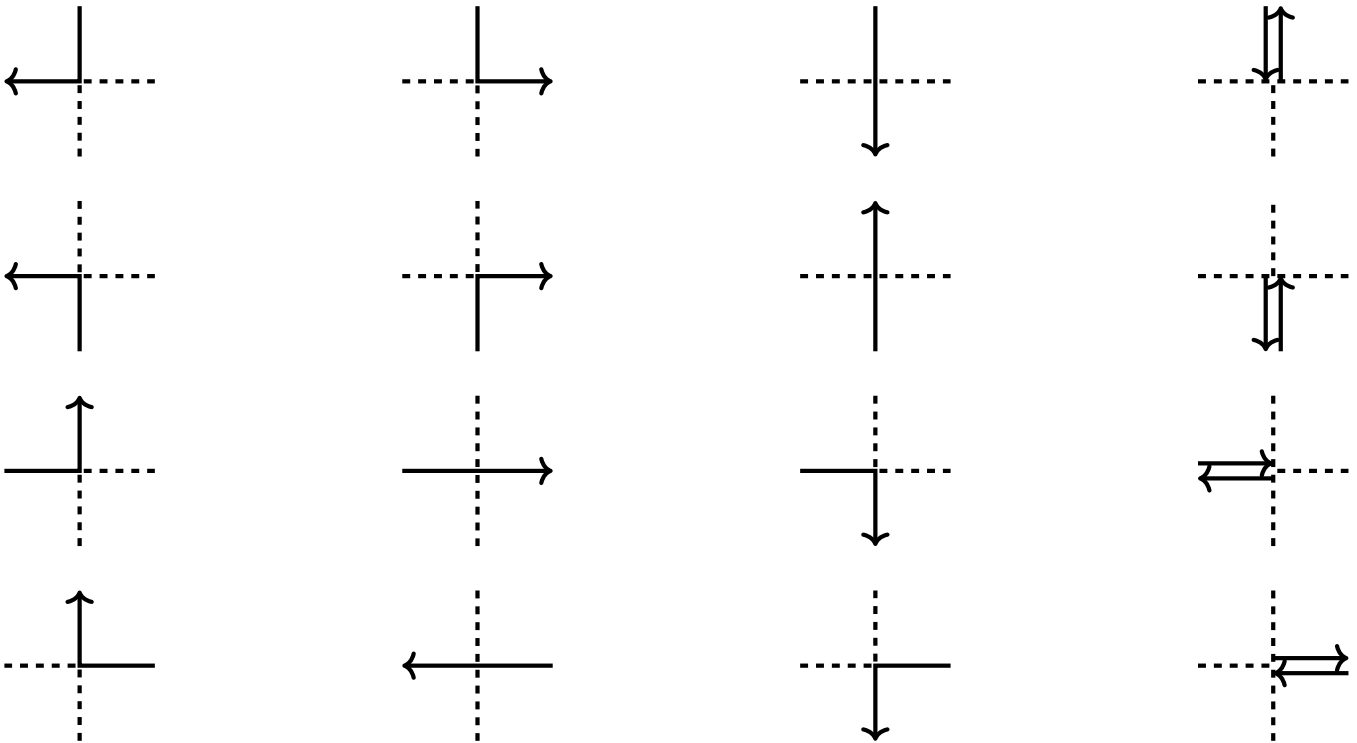
\begin{figure*}
\begin{tikzpicture}
 \draw[step=1.0,black,ultra thick,dashed] (0.0,-1.0) -- (0.0,0.0) -- (1.0,0.0);  
 \draw[->,black,ultra thick]  (0.0,1.0) -- (0.0 ,0.0) -- (-1.0,0.0) ; 
\end{tikzpicture}
\hfill
\begin{tikzpicture}
 \draw[step=1.0,black,ultra thick,dashed] (-1.0,0.0) -- (0.0,0.0) -- (0.0,-1.0);  
 \draw[->,black,ultra thick]  (0.0,1.0) -- (0.0 ,0.0) -- (1.0,0.0) ;
 \end{tikzpicture}
 \hfill
  \begin{tikzpicture}
 \draw[step=1.0,black,ultra thick,dashed] (-1.0,0.0) -- (0.0,0.0) -- (1.0,0.0);  
 \draw[->,black,ultra thick]  (0.0,1.0) -- (0.0 ,0.0) -- (0.0,-1.0) ;
 \end{tikzpicture}
 \hfill 
  \begin{tikzpicture}
 \draw[step=1.0,black,ultra thick,dashed] (-1.0,0.0) -- (0.0,0.0) -- (1.0,0.0) ;
   \draw[step=1.0,black,ultra thick,dashed] (0.0,-1.0) -- (0.0,0.0) ;
 \draw[->,black,ultra thick]  (-0.1,1.0) -- (-0.1 ,0.0) ;
 \draw[->,black,ultra thick]  (0.1,0.0) -- (0.1,1.0) ;
 \end{tikzpicture}
 
 \vspace{0.5cm}
 
 \begin{tikzpicture}
 \draw[step=1.0,black,ultra thick,dashed] (0.0,1.0) -- (0.0,0.0) -- (1.0,0.0);  
 \draw[->,black,ultra thick]  (0.0,-1.0) -- (0.0 ,0.0) -- (-1.0,0.0) ; 
\end{tikzpicture}
\hfill
\begin{tikzpicture}
 \draw[step=1.0,black,ultra thick,dashed] (-1.0,0.0) -- (0.0,0.0) -- (0.0,1.0);  
 \draw[->,black,ultra thick]  (0.0,-1.0) -- (0.0 ,0.0) -- (1.0,0.0) ;
 \end{tikzpicture}
 \hfill
 \begin{tikzpicture}
 \draw[step=1.0,black,ultra thick,dashed] (-1.0,0.0) -- (0.0,0.0) -- (1.0,0.0);  
 \draw[->,black,ultra thick]  (0.0,-1.0) -- (0.0 ,0.0) -- (0.0,1.0) ;
 \end{tikzpicture}
 \hfill 
  \begin{tikzpicture}
 \draw[step=1.0,black,ultra thick,dashed] (-1.0,0.0) -- (0.0,0.0) -- (1.0,0.0) ;
   \draw[step=1.0,black,ultra thick,dashed] (0.0,0.0) -- (0.0,1.0) ;
 \draw[->,black,ultra thick]  (-0.1,0.0) -- (-0.1 ,-1.0) ;
 \draw[->,black,ultra thick]  (0.1,-1.0) -- (0.1,0.0) ;
 \end{tikzpicture}
 
  \vspace{0.5cm}
 
 \begin{tikzpicture}
 \draw[step=1.0,black,ultra thick,dashed] (0.0,-1.0) -- (0.0,0.0) -- (1.0,0.0);  
 \draw[->,black,ultra thick]  (-1.0,0.0) -- (0.0 ,0.0) -- (0.0,1.0) ; 
\end{tikzpicture}
\hfill
\begin{tikzpicture}
 \draw[step=1.0,black,ultra thick,dashed] (0.0,1.0) -- (0.0,0.0) -- (0.0,-1.0);  
 \draw[->,black,ultra thick]  (-1.0,0.0) -- (0.0 ,0.0) -- (1.0,0.0) ;
 \end{tikzpicture}
  \hfill
 \begin{tikzpicture}
 \draw[step=1.0,black,ultra thick,dashed] (0.0,1.0) -- (0.0,0.0) -- (1.0,0.0);  
 \draw[->,black,ultra thick]  (-1.0,0.0) -- (0.0 ,0.0) -- (0.0,-1.0) ;
 \end{tikzpicture}
 \hfill 
  \begin{tikzpicture}
 \draw[step=1.0,black,ultra thick,dashed] (0.0,1.0) -- (0.0,0.0) -- (1.0,0.0) ;
   \draw[step=1.0,black,ultra thick,dashed] (0.0,-1.0) -- (0.0,0.0) ;
 \draw[->,black,ultra thick]  (-1.0,0.1) -- (0.0,0.1) ;
 \draw[->,black,ultra thick]  (0.0,-0.1) -- (-1.0,-0.1) ;
 \end{tikzpicture}
 
 \vspace{0.5cm}
 
  \begin{tikzpicture}
 \draw[step=1.0,black,ultra thick,dashed] (0.0,-1.0) -- (0.0,0.0) -- (-1.0,0.0);  
 \draw[->,black,ultra thick]  (1.0,0.0) -- (0.0 ,0.0) -- (0.0,1.0) ; 
\end{tikzpicture}
\hfill
\begin{tikzpicture}
 \draw[step=1.0,black,ultra thick,dashed] (0.0,1.0) -- (0.0,0.0) -- (0.0,-1.0);  
 \draw[->,black,ultra thick]  (1.0,0.0) -- (0.0 ,0.0) -- (-1.0,0.0) ;
 \end{tikzpicture}
  \hfill
  \begin{tikzpicture}
 \draw[step=1.0,black,ultra thick,dashed] (-1.0,0.0) -- (0.0,0.0) -- (0.0,1.0);  
 \draw[->,black,ultra thick]  (1.0,0.0) -- (0.0 ,0.0) -- (0.0,-1.0) ;
 \end{tikzpicture}
 \hfill 
  \begin{tikzpicture}
 \draw[step=1.0,black,ultra thick,dashed] (0.0,1.0) -- (0.0,0.0) -- (0.0,-1.0) ;
   \draw[step=1.0,black,ultra thick,dashed] (-1.0,0.0) -- (0.0,0.0) ;
 \draw[->,black,ultra thick]  (1.0,-0.1) -- (0.0,-0.1) ;
 \draw[->,black,ultra thick]  (0.0,0.1) -- (1.0,0.1) ;
 \end{tikzpicture}
\caption{Sixteen non-zero possibilities for $ \psi, \bar{\psi} $ integration at a site.  These 16 possibilities end up being exactly the nonzero elements of the fermion tensor.}
\label{fig:vertex_tensor}
\end{figure*}

Following Ref.~\cite{GATTRINGER2015732}, and using the same notaion for clarity, we first integrate out the
fermions and generate an effective action depending only on the gauge fields.
As a first step we redefine the link variables such that the staggered fermion phases $ \eta_{\mu}(x) $ can be absorbed into modified link variables $U_\mu(x)\to \eta_\mu(x)U_\mu(x)$. Under this transformation the
gauge action picks up an overall negative sign but the measure is invariant. The Boltzmann factor associated with
each bilinear fermion term can be written as the product of forward and backward hopping terms yielding
a partition function
\begin{align}
\nonumber
Z_F = &\int D[U] D[\bar{\psi}] D[\psi] \times \\ \nonumber
&\prod_x \sum_{k=0}^{1} \left( \frac{1}{2}\bar{\psi}(x)U_\mu(x)\psi(x+\mu) \right)^{k} \times \\
&\sum_{\bar{k}=0}^{1} \left( \frac{1}{2}\bar{\psi}(x+\mu)U^\dagger_\mu(x)\psi(x) \right)^{\bar{k}}.
\end{align}
Notice that higher order terms in the expansion of the Boltzmann factors vanish because of the Grassmann nature of the fermions.
There are several ways to generate a non-zero contribution to $Z_F$.  In each case, the Grassmann integration at each site must be saturated.
To saturate the Grassmann integrations, exactly one forward and one backward hopping term must be associated with each site.  This gives rise to a simple collection of possibilities.  On the one hand, there may be a single forward and backward hop along the same link.  This obviously saturates the integration, and is referred to as a dimer.  On the other hand, there may be a forward and backward hop on two different links at a site.  This indicates the passage of fermionic current through the site, and again saturates the integration measure there.  Furthermore because of gauge invariance
any non-dimer contribution to $Z_F$ must correspond
to a closed loop. Fig.~\ref{fig:vertex_tensor} shows the allowed site contributions. A bold
link indicates the presence of a $\frac{1}{2}U$ or a $-\frac{1}{2}U^\dagger$ factor along that link. Notice that the
links are oriented corresponding to the presence of an arrow on each bold link whose direction is conserved
through a site.

For a loop $\ell$ with length $L(\ell)$ one finds a
contribution with absolute value
\begin{equation}
\left(\frac{1}{2}\right)^{L(\ell)}\prod_{x,\mu\in \ell}\left(U_\mu \right)^{k_\mu(x)}
\end{equation}
where $ k_{\mu}(x) = \pm 1 $ distinguishes between $U_\mu(x)$ and $U^\dagger_\mu(x)$.
In addition each loop carries a certain $Z_2$ phase which depends on the
length of the loop and its winding along the temporal direction given by 
\begin{equation}
- (-1)^{\frac{1}{2}L(l)}(-1)^{W(l)}.
\end{equation}
Here, the overall negative sign is the usual one for closed fermion loops while the second
factor keeps track of the number of forward hops which is exactly half the total length of the loop
for a closed loop. Finally the factor $ (-1)^{W(l)}$  
of the loop will be determined by the number of windings of the loop along the temporal direction assuming
anti-periodic boundary conditions for the fermions.
Using dimers and loops as basic constituents for non-zero contributions to the fermionic partition function we can write 
\begin{align}
\label{eq:z_f}
\nonumber
Z_F  = \left( \frac{1}{2} \right)^{V} &\sum_{l,d} (-1) ^{N_L + \frac{1}{2} \sum_{l} L(l) + \sum_{l} W(l)} \times \\
&\prod_{l} \left[ \prod_{x,\mu \in l} U^{k_{\mu}(x)}_{\mu}(x) \right].
\end{align}
To proceed further we will need to construct this loop representation
from the contraction of more basic objects located at sites  and we take up this task in the next section.

\subsection{Tensor Formulation of the Fermionic Partition Function}

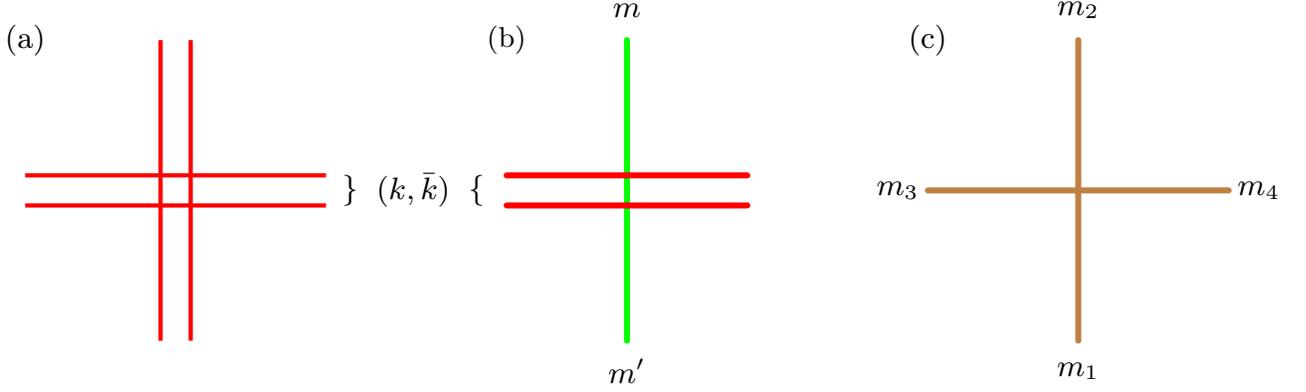
\begin{figure*}
\begin{center}
\begin{tikzpicture}[scale=0.40]
\draw node[scale=1.3] at (-5., 5.) {(a)};
\draw node[scale=1.3] at (5.8,0) { $ \}$};
\draw node[scale=1.3] at (7.9,0) { $ (k , \bar{k})$};
\draw node[scale=1.3] at (10.,0) { $ \{$};
\draw[red,ultra thick]  (-5.,0.5) -- (5.0 ,0.5) ; 
\draw[red,ultra thick]  (-5.0,-0.5) -- (5. ,-0.5) ; 
\draw[red,ultra thick]  (-0.5, 5.) -- (-0.5 ,-5.0) ; 
\draw[red,ultra thick]  (0.5,-5.0) -- (0.5 , 5.) ;
        
\draw node[scale=1.2] at (11., 5.) {(b)};
\draw node[scale=1.3] at (15.0,6.0) { $m$};
\draw node[scale=1.3] at (15.0,-6.0) { $m^{\prime}$};
\draw[green,cap=round,line width = 0.8mm] (15.0,-5.0) -- (15.0,5.0);
\draw[red,cap=round,line width = 0.8mm]  (15.0,0.5) -- (19.0 ,0.5) ; 
\draw[red,cap=round,line width = 0.8mm]  (19.0,-0.5) -- (15.0 ,-0.5) ; 
\draw[red,cap=round,line width = 0.8mm]  (11,0.5) -- (15.0 ,0.5) ;  
\draw[red,cap=round,line width = 0.8mm]  (15.0,-0.5) -- (11.0 ,-0.5) ; 

\draw node[scale=1.3] at (25., 5.) {(c)};   
\draw node[scale=1.3] at (30.0,-6.0) { $m_{1}$};
\draw node[scale=1.3] at (30.0,6.0) { $m_{2}$};
\draw node[scale=1.3] at (24.0,.0) { $m_{3}$};
\draw node[scale=1.3] at (36.0,0.0) { $m_{4}$};
\draw[cap=round,line width = 0.8mm,brown] (30.0,-5.0) -- (30.0,5.0);
\draw[cap=round,line width = 0.8mm,brown]  (25.0,0.0) -- (35.0,0.0) ; 
\end{tikzpicture}
\caption{(a) Fermion tensor associated with the sites of the lattice.  The two lines in each direction can take on the values of unoccupied, or occupied with a forward or backwards current.  Each pair can then have four states, unoccupied, outgoing fermionic current, incomming fermionic current, and both outgoing and incommming current, \emph{i.e.} a dimer. (b) The constraint tensor associated with the links.  This tensor enforces that the difference between the $m$ electric field numbers appropriately matches, and compensates, the fermionic current accross the link. (c) The gauge field tensor associated with each plaquette.  This tensor has four indices, but the only non-vanishing elements are when all indices take the same value, \emph{i.e.} it is diagonal in all four indices.  Each nonzero element is associated with weight factors given by modified Bessel functions.}
\label{fig:three_tensors}
\end{center}
\end{figure*}

We need to construct a local tensor 
which under contraction along lattice links yields $Z_F$. Let us ignore the overall sign for now and just deal with the magnitude. We allow two types of indices per link to capture
separately the incoming and outgoing fermion lines making the fermion site tensor a
rank eight object. Since each site is either the endpoint of a dimer, or has fermionic current incoming and outgoing from it is
then modeled by the tensor structure (we leave off the gauge link factors for now)
\begin{equation}
T^{x}_{k_1\bar{k_1}k_2\bar{k_2}k_3\bar{k_3}k_4\bar{k_4}} =   \left\{
	\begin{array}{ll}
		 1 & \text{if any two $k_i$ and $\bar{k_i}$ are} \\
		 & \text{one and others are zero.}     \\
		 0  &  \text{otherwise}
		  \end{array}
\right.
\end{equation}
where each $ (k_i ,\bar{k_i} ) = 0,1$.  A graphical representation of this tensor is shown in Fig.~\ref{fig:three_tensors}~(a).
By repeatedly contracting this site tensor with copies of itself over the lattice it can be seen that
we generate the full set of closed loops and dimers for the
model at zero gauge coupling {\rm excluding} the overall factor of minus one for each closed fermion loop.  The absolute value of the partition function at zero gauge coupling is then,
\begin{equation}
Z_F^{\beta=\infty}= \sum_{\{k,\bar{k}\}} \prod_{x} T^{x}_{k_1\bar{k_1}k_2\bar{k_2}k_3\bar{k_3}k_4\bar{k_4}}.
\end{equation}
Here, $ \{k,\bar{k}\} $ denote the set of $ k,\bar{k} $ values for the entire lattice. 
Said another way, the 16 possible vertex configurations for fermion hopping in Fig.~\ref{fig:vertex_tensor} are captured as nonzero tensor elements in the $T$ tensor.

\subsection{Integrating out the gauge fields}
The fermion partition function in the previous section does not include any contribution or interaction with the gauge fields.  To proceed further we will
employ a character expansion of the Boltzmann factors associated with
the gauge action. This will ensure that each plaquette in the lattice will carry
an integer variable. Integration of the link gauge field in the background of
a particular set of fermion loops restricts the plaquette variables to change
by plus or minus one on crossing any fermion line.

In this section, we will describe this in detail and, along with the tensor from the previous section, construct a tensor network that when fully contracted reproduces the full partition function for the massless Schwinger model.

To integrate the gauge links we first start by performing a character expansion on the Boltzmann factor
corresponding to the pure gauge plaquette action 
\begin{align}
\nonumber
&e^{-\beta \cos{\left[A_{\mu}(x) + A_{\nu}(x+\mu) - A_{\mu}(x+\nu) - A_{\nu}(x)\right]} } = \\
&\sum_{m=-\infty}^{m=\infty} I_{m}(-\beta) e^{i m \left[A_{\mu}(x) + A_{\nu}(x+\mu) - A_{\mu}(x+\nu) - A_{\nu}(x)\right] } .
\end{align}
Each plaquette $p$ is now labeled by an integer $m_p$.  Note that $I_{m}(-\beta) = (-1)^{m} I_{m}(\beta)$. Furthermore,
each link $\ell$ is shared by two plaquettes $p$ and $p^\prime$ each of which supplies a 
factor of $e^{i m_p A_\ell}$ and $e^{-i m_{p^\prime} A_\ell}$. 
In addition, the link carries a factor of $e^{ik_\ell A_\ell}$
or $e^{-i\bar{k}_\ell A_\ell}$ coming from $Z_F$. Thus, in total, links carry two $m$ indices inherited from
their neighboring plaquettes together with a $k$ and a $\bar{k}$ index associated with the fermionic
hopping terms.
The integral over the link field then gives
\begin{equation}
\label{eq:cnst}
\int_{-\pi}^{\pi} \frac{dA_\ell}{2\pi} e^{i (m_p - m_{p^\prime} + k_\ell - \bar{k_\ell} ) A_\ell } = \delta_{ m_p - m_{p^\prime} + k_\ell -\bar{k_\ell}, 0} .
\end{equation}
This allows us to write the partition function as
\begin{align}
\nonumber
Z &= \sum_{ \{m_p\} }  \sum_{\{k_\ell,\bar{k_\ell}\}}  \prod_\ell \delta_{ m_p - m_{p^\prime} + k_\ell -\bar{k_\ell},0} \prod_{p}I_{m_p}(\beta) \times \\
&\prod_{x} T^{x}_{k_1\bar{k_1}k_2\bar{k_2}k_3\bar{k_3}k_4\bar{k_4}}
\times (-1)^{N_{L} + N_{P} + \frac{1}{2}\sum_{l} L(l)}
 \label{ZZ}
 \end{align}
 where $\{m_p\}$ denotes the set of plaquette integers over the entire lattice, $\{k_\ell,\bar{k}_\ell\}$
 represent $k$ indices over the links, and $N_{P} = \sum_{p} m_{p}$.  At this 
 point we have included all the minus signs for completeness. For periodic boundary conditions, the sum of winding numbers must always be zero, since one is restricted to the total charge-0 sector of the theory.  Note that for this situation the overall $\pm 1$ factor is always positive~\cite{GATTRINGER2015732}.
 
Now, associated with each link are $m$ fields and $k$ fields, and a constraint between them.  Associated with each plaquette is a single $m$ field.  This lets us define a link tensor, and a plaquette tensor.  Link tensors have indices connecting to fermion tensors (the $T$ tensors) living on each site, and guage-field indices connecting to plaquette tensors (on each plaquette).  We define this link tensor, $A$, as,
\begin{equation}
A_{m_i m_j k_a \bar{k_a} k_b \bar{k_b}} \equiv \delta_{ m_i - m_j  + k_a -\bar{k_a},0}\delta_{k_a,k_b} \delta_{\bar{k_a} ,\bar{k_b}}.
\end{equation}
Fermion-like indices on link tensors are purely diagonal as seen from the definition involving the $\delta$ function constraints on links. A diagram showing the relative position of the fermion and plaquette indices is shown in Fig.~\ref{fig:three_tensors}~(b).  Since there is only a single $m$ associated with each plaquette, a tensor definition must only depend on that single $m$. A plaquette tensor, $B$, can be defined as,
\begin{equation}
B_{m_1m_2m_3m_4} =  \left\{
	\begin{array}{ll}
		 I_{m}(\beta) & \text{if } m_1=m_2=m_3=m_4 \\
		 & \quad = m \\
		 0 & {\rm otherwise   .}
	\end{array}
\right.
\end{equation}
A graphical representation for the $ B $ tensor associated with plaquettes is shown in Fig.~\ref{fig:three_tensors}~(c).

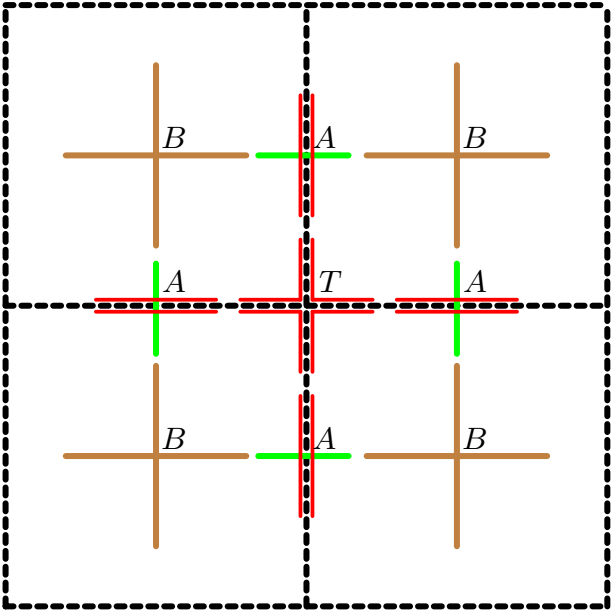
\begin{figure}
\centering
\begin{tikzpicture}[scale=0.8]
\draw[step=5cm,black,cap=round,line width=0.8mm,dashed] (0.0,0.0) grid (10.0,10.0);
\draw[cap=round,line width = 0.8mm,brown] (1.0,2.5) -- (4.0,2.5);
\draw[cap=round,line width = 0.8mm,brown]  (2.5,1.0) -- (2.5,4.0) ; 
\draw node[scale=1.3] at (2.8,2.8) { $B$};
\draw[cap=round,line width = 0.8mm,brown] (6.0,2.5) -- (9.0,2.5);
\draw[cap=round,line width = 0.8mm,brown]  (7.5,1.0) -- (7.5,4.0) ;
\draw node[scale=1.3] at (7.8,2.8) { $B$};
\draw[cap=round,line width = 0.8mm,brown] (1.0,7.5) -- (4.0,7.5);
\draw[cap=round,line width = 0.8mm,brown]  (2.5,6.0) -- (2.5,9.0) ; 
\draw node[scale=1.3] at (7.8,7.8) { $B$};
\draw[cap=round,line width = 0.8mm,brown] (6.0,7.5) -- (9.0,7.5);
\draw[cap=round,line width = 0.8mm,brown]  (7.5,6.0) -- (7.5,9.0) ; 
\draw node[scale=1.3] at (2.8,7.8) { $B$};
\draw[green,cap=round,line width = 0.8mm] (2.5,4.2) -- (2.5,5.7);
\draw[red,cap=round,line width = 0.5mm]  (1.5,4.9) -- (3.5 ,4.9) ; 
  \draw[red,cap=round,line width = 0.5mm]  (1.5,5.1) -- (3.5 ,5.1) ; 
  \draw node[scale=1.3] at (2.8,5.4) { $A$};
   \draw[green,cap=round,line width = 0.8mm] (7.5,4.2) -- (7.5,5.7);
\draw[red,cap=round,line width = 0.5mm]  (6.5,4.9) -- (8.5 ,4.9) ; 
  \draw[red,cap=round,line width = 0.5mm]  (6.5,5.1) -- (8.5 ,5.1) ; 
   \draw node[scale=1.3] at (7.8,5.4) { $A$};
   \draw[green,cap=round,line width = 0.8mm] (4.2,2.5) -- (5.7,2.5);
\draw[red,cap=round,line width = 0.5mm]  (4.9,1.5) -- (4.9 ,3.5) ; 
 \draw[red,cap=round,line width = 0.5mm]  (5.1,1.5) -- (5.1 ,3.5) ; 
  \draw node[scale=1.3] at (5.3,2.8) { $A$};
   \draw[green,cap=round,line width = 0.8mm] (4.2,7.5) -- (5.7,7.5);
\draw[red,cap=round,line width = 0.5mm]  (4.9,6.5) -- (4.9 ,8.5) ; 
 \draw[red,cap=round,line width = 0.5mm]  (5.1,6.5) -- (5.1 ,8.5) ; 
 \draw node[scale=1.3] at (5.3,7.8) { $A$};
 %
  \draw[red,cap=round,line width = 0.5mm]  (5.1,5.1) -- (6.1 ,5.1) ; 
  \draw[red,cap=round,line width = 0.5mm]  (5.1,5.1) -- (5.1 ,6.1) ;
   \draw[red,cap=round,line width = 0.5mm]  (3.9,5.1) -- (4.9 ,5.1) ;
   \draw[red,cap=round,line width = 0.5mm]  (4.9,5.1) -- (4.9 ,6.1) ;
   \draw[red,cap=round,line width = 0.5mm]  (5.1,3.9) -- (5.1 ,4.9) ;
   \draw[red,cap=round,line width = 0.5mm]  (5.1,4.9) -- (6.1 ,4.9) ;
   \draw[red,cap=round,line width = 0.5mm]  (3.9,4.9) -- (4.9 ,4.9) ;
   \draw[red,cap=round,line width = 0.5mm]  (4.9,3.9) -- (4.9,4.9) ;
    \draw node[scale=1.3] at (5.4,5.4) { $T$};
   \end{tikzpicture}
   \caption{Elementary tensors $T$, $A$, and $B$.  When these tensors are contracted in the pattern shown here the world-line representation of the partition function is generated exactly.}
   \label{fig:tn}
\end{figure}

These definitions of  the $ A $ and $ B $ tensors allow us to write the partition function as follows,
\begin{align}
\nonumber
Z = \sum_{\{k,\bar{k}\}} \sum_{ \{m_p\} } &\left( \prod_{p}  B_{m_im_jm_km_l} \right)\left( \prod_{l}  A_{m_i m_j k_a \bar{k_a} k_b \bar{k_b}} \right) \times \\
&\left( \prod_{x}  T_{k_a\bar{k_a}k_b\bar{k_b}k_c\bar{k_c}k_d\bar{k_d}}  \right).
\end{align}
This contraction over three unique tensor types can be represented as the tensor network shown in Fig.~\ref{fig:tn}.  Since the fermionic $k$ indices always come in $k$, $\bar{k}$ pairs, we can form a product state of those two indices to reduce the complexity of the notation,
\begin{align}
\nonumber
    T \rightarrow T' &= T_{(k_a \otimes \bar{k_a})(k_b \otimes \bar{k_b})(k_c \otimes \bar{k_c})(k_d \otimes \bar{k_d})} \\ &= T_{K_a K_b K_c K_d}.
\end{align}
\begin{align}
    A \rightarrow A' = A_{m_i m_j (k_a \otimes \bar{k_a})( k_b \otimes \bar{k_b})} = A_{m_i m_j K_a K_b}
\end{align}
The new enlarged $K$ indices take values from 0 to 3, enumerating the four possible states each link can have: unoccupied, incoming, outgoing, and dimer.  The $A$ tensors are still diagonal in the new $K$ indices.

\section{Transfer Matrix}
Using the tensors defined in the previous sections, one can build a transfer matrix for this model.  The transfer matrix can be defined as the product of two types of matrices.  In this section, we first define and construct these two different matrices.  Then, by combining these two matrices in the appropriate way we can define a transfer matrix.  The partition function is the trace of the $ N_{\tau}^{\text{th}} $ power of this final matrix.

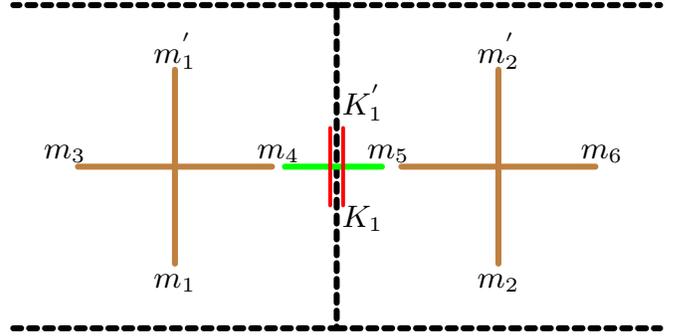
\begin{figure}
\centering
\begin{tikzpicture}[scale=0.86]
\draw[black, cap=round,line width=0.8mm, dashed] (0,0) -- (5,0) -- (5,5)  -- (10,5);
\draw[black, cap=round,line width=0.8mm, dashed] (5,0) -- (10,0);
\draw[black,cap=round,line width=0.8mm,dashed] (5,5) -- (0,5); 
\draw[cap=round,line width=0.8mm,brown] (1.0,2.5) -- (4.0,2.5);
\draw[cap=round,line width = 0.8mm,brown]  (2.5,1.0) -- (2.5,4.0) ; 
\draw node[scale=1.3] at (0.8,2.7) {$m_3$};
\draw node[scale=1.3] at (4.1,2.7) {$m_4$};
\draw node[scale=1.3] at (5.8,2.7) {$m_5$};
\draw node[scale=1.3] at (9.1,2.7) {$m_6$};
\draw node[scale=1.3] at (2.5,4.3) {$m^{'}_1$};
\draw node[scale=1.3] at (2.5,0.7) {$m_1$};
\draw node[scale=1.3] at (7.5,4.3) {$m^{'}_2$};
\draw node[scale=1.3] at (7.5,0.7) {$m_2$};
\draw[cap=round,line width = 0.8mm,brown] (6.0,2.5) -- (9.0,2.5);
\draw[cap=round,line width = 0.8mm,brown]  (7.5,1.0) -- (7.5,4.0) ; 
 \draw[green,cap=round,line width = 0.8mm] (4.2,2.5) -- (5.7,2.5);
\draw[red,cap=round,line width = 0.5mm]  (4.9,1.9) -- (4.9 ,3.1) ; 
\draw[red,cap=round,line width = 0.5mm]  (5.1,1.9) -- (5.1 ,3.1) ; 
  \draw node[scale=1.3] at (5.4, 1.7) {$K_1$};
    \draw node[scale=1.3] at (5.4, 3.5) {$K^{'}_1$};
 \end{tikzpicture}
 \caption{Construction of part of the $\mathcal{B}$ matrix.  In principle the construction continues to the left and right with $A$ tensors contracted with the $B$ tensors, and so on.}
\label{fig:B}
\end{figure}

The first type of matrix we define is the $ \mathcal{B} $  matrix.  It is made by contracting alternating $B$ and $A$ tensors along a time-slice.  
\begin{align}
\nonumber
&\mathcal{B}_{(m_1\otimes \cdots m_N \otimes K_{1}\otimes \cdots K_{N} ) ( m^{'}_1 \otimes \cdots m^{'}_N \otimes K_{1}^{\prime} \otimes \cdots K_{N}^{\prime})} =   \\ \nonumber
&B_{m m^{'} m_1 m^{'}_1} A_{m^{'} m^{''} K_{1} K_{1}^{\prime}} B_{m^{''}m^{'''} m_2 m^{'}_2} \times \\
& A_{m^{'''}m^{''''} K_{2} K_{2}^{\prime}} \cdots B_{m^{(N-1)} m m_N m^{'}_N} 
\end{align}
where a sum over repeated indices is implied.  Diagrammatically $ \mathcal{B}$ is represented as Fig.~\ref{fig:B}.  An important feature of this matrix is that it is diagonal, due to the diagonal nature of the $B$ tensors, and the $K$ indices in the $A$ tensors.  This means incoming states through this matrix do not change into other states.

\begin{figure*}
\begin{tikzpicture}
\draw[black, cap=round,line width=0.8mm, dashed] (0,0) -- (5,0) -- (15,0);
\draw[black, cap=round,line width=0.8mm, dashed] (5,-2) -- (5,2) ;
\draw[black, cap=round,line width=0.8mm, dashed] (10,-2) -- (10,2);

\draw[green,cap=round,line width = 0.8mm] (2.5,-1.1) -- (2.5,1.1);
\draw[red,cap=round,line width = 0.5mm]  (1.5,0.1) -- (3.5 ,0.1) ; 
\draw[red,cap=round,line width = 0.5mm]  (1.5,-0.1) -- (3.5 ,-0.1) ; 

\draw[green,cap=round,line width = 0.8mm] (7.5,-1.1) -- (7.5,1.1);
\draw[red,cap=round,line width = 0.5mm]  (6.5,0.1) -- (8.5 ,0.1) ; 
\draw[red,cap=round,line width = 0.5mm]  (6.5,-0.1) -- (8.5 ,-0.1) ; 

\draw[green,cap=round,line width = 0.8mm] (12.5,-1.1) -- (12.5,1.1);
\draw[red,cap=round,line width = 0.5mm]  (11.5,0.1) -- (13.5 ,0.1) ; 
\draw[red,cap=round,line width = 0.5mm]  (11.5,-0.1) -- (13.5 ,-0.1) ; 
 
 \draw node[scale=1.3] at (2.5, 1.3) {$m_1$};
  \draw node[scale=1.3] at (2.5, -1.3) {$m^{'}_1$};

   \draw node[scale=1.3] at (7.5, 1.3) {$m_2$};
  \draw node[scale=1.3] at (7.5, -1.3) {$m^{'}_2$};
  
   \draw node[scale=1.3] at (12.5, 1.3) {$m_3$};
  \draw node[scale=1.3] at (12.5, -1.3) {$m^{'}_3$};

 \draw node[scale=1.3] at (1.25, 0.4) {$\bar{K}_1$};

 \draw node[scale=1.3] at (3.75, 0.4) {$\bar{K}_2$};
  
   \draw node[scale=1.3] at (6.25, 0.4) {$\bar{K}_{3}$};
  
   \draw node[scale=1.3] at (8.75, 0.4) {$\bar{K}_4$};
  
   \draw node[scale=1.3] at (11.25, 0.4) {$\bar{K}_5$};
 
  \draw node[scale=1.3] at (5.3,1.3) {$K_1$};
  
  \draw node[scale=1.3] at (5.3,-1.3) {$K^{'}_1$};
  
  \draw node[scale=1.3] at (10.3,1.3) {$K_2$};
  
  \draw node[scale=1.3] at (10.3,-1.3) {$K^{'}_2$};

   \draw[red,cap=round,line width = 0.5mm]  (5.1,0.1) -- (6.1 ,0.1) ; 
  \draw[red,cap=round,line width = 0.5mm]  (5.1,0.1) -- (5.1 ,1.1) ;
   \draw[red,cap=round,line width = 0.5mm]  (3.9,0.1) -- (4.9 ,0.1) ;
   \draw[red,cap=round,line width = 0.5mm]  (4.9,0.1) -- (4.9 ,1.1) ;
   \draw[red,cap=round,line width = 0.5mm]  (5.1,-1.1) -- (5.1 ,-0.1) ;
   \draw[red,cap=round,line width = 0.5mm]  (5.1,-0.1) -- (6.1 ,-0.1) ;
   \draw[red,cap=round,line width = 0.5mm]  (3.9,-0.1) -- (4.9 ,-0.1) ;
   \draw[red,cap=round,line width = 0.5mm]  (4.9,-1.1) -- (4.9,-0.1) ;
   
    \draw[red,cap=round,line width = 0.5mm]  (10.1,0.1) -- (11.1 ,0.1) ; 
  \draw[red,cap=round,line width = 0.5mm]  (10.1,0.1) -- (10.1 ,1.1) ;
   \draw[red,cap=round,line width = 0.5mm]  (8.9,0.1) -- (9.9 ,0.1) ;
   \draw[red,cap=round,line width = 0.5mm]  (9.9,0.1) -- (9.9 ,1.1) ;
   \draw[red,cap=round,line width = 0.5mm]  (10.1,-1.1) -- (10.1 ,-0.1) ;
   \draw[red,cap=round,line width = 0.5mm]  (10.1,-0.1) -- (11.1 ,-0.1) ;
   \draw[red,cap=round,line width = 0.5mm]  (8.9,-0.1) -- (9.9 ,-0.1) ;
   \draw[red,cap=round,line width = 0.5mm]  (9.9,-1.1) -- (9.9,-0.1) ;

\end{tikzpicture}
\caption{Construction of matrix $\mathcal{A}$.  In principle the construction continues to the left and right, alternating contraction between $A$ and $T$ tensors.  This matrix is responsible for moving fermionic current around in space and time, and adjusting the gradient of the electric field to compensate.}
\label{fig:amatrix}
\end{figure*}
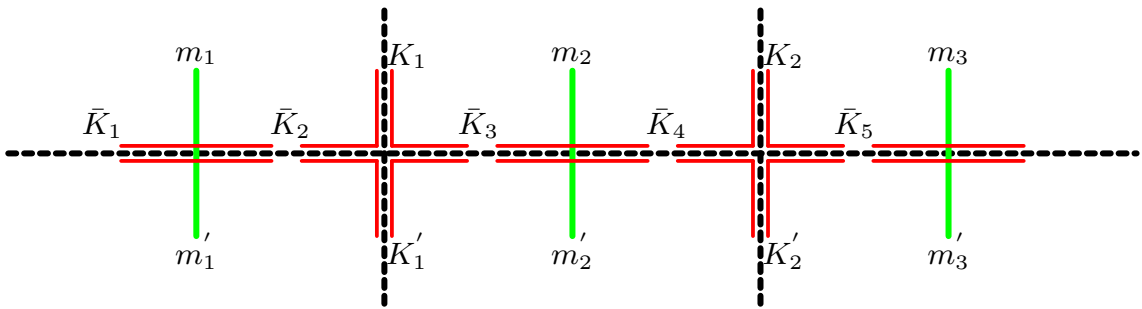

\vspace{0.5cm}
In analogy with the construction of $ \mathcal{B} $ we define the $\mathcal{A}$ matrix as the alternating contraction of $T$ and $A$ tensors along a time-slice,
\begin{align}
\nonumber
&\mathcal{A}_{(m_1 \otimes \cdots m_N \otimes K_{1} \otimes \cdots K_{N} ) ( m^{'}_1 \otimes \cdots m^{'}_N \otimes K_{1}^{\prime} \otimes \cdots K_{N}^{\prime})} =  \\  & A_{m_1 m^{'}_1 \bar{K}_1 \bar{K}_2 } T_{\bar{K}_{2} \bar{K}_{3} K_{1} K_{1}^{\prime} } A_{m_2 m^{'}_2 \bar{K}_{3} \bar{K}_{4}} \cdots A_{m_N m^{'}_N \bar{K}_{N} \bar{K}_{1} } 
\end{align}
with a diagrammatic representation given by Fig.~\ref{fig:amatrix}.  This matrix has off-diagonal elements, and is responsible for the changing of states between time-slices.  This matrix moves fermionic current across space, and through time, with the appropriate shift in the electric field to balance.

\vspace{0.5cm}

Using the definitions above we can recast the partition function into an alternating product of $ \mathcal{B} $ and $\mathcal{A}$ matrices.  This alternating product can be broken up, and recast as the $N_{\tau}^{\text{th}}$ power of a single matrix,
\begin{equation}
\label{eq:tm}
 \mathcal{T}_{\alpha \beta}  = \sqrt{\mathcal{B}}_{\alpha \delta} \mathcal{A}_{\delta \gamma} \sqrt{\mathcal{B}}_{\gamma \beta}
\end{equation}
where the square root is well-defined since $ \mathcal{B}$ is diagonal in all of it's indices (and its matrix elements are positive). The indices in Eq.~\eqref{eq:tm} are collective indices as defined before in the definitions of the $\mathcal{B}$ and $\mathcal{A}$ tensors. Now we can write the partition function as follows,
\begin{equation}
 Z = \Tr [ \mathcal{T}^{N_t} ].
\end{equation}

\section{Fundamental Tensor for TRG}

\subsection{Asymmetric tensor}
In order to have efficient numerical calculations using the TRG, the tensor network structure should be translationally invariant.  This means that for whatever fundamental tensor one uses, it must contract naturally with itself.  That is, the top indices of the fundamental tensor should be compatible for contraction with the bottom indices, and the indices on the left side of the tensor should be compatible for contraction with the indices on the right.

\begin{figure}[t]
\centering
\begin{tikzpicture}[scale=0.8]

\draw[step=5cm,black,cap=round,line width=0.8mm,dashed] (0.0,0.0) grid (10.0,10.0);
\draw[cap=round,line width = 0.8mm,brown] (1.0,2.5) -- (4.0,2.5);
\draw[cap=round,line width = 0.8mm,brown]  (2.5,1.0) -- (2.5,4.0) ; 
\draw node[scale=1.3] at (2.8,2.8) { $B$};

\draw[cap=round,line width = 0.8mm,brown] (6.0,2.5) -- (9.0,2.5);
\draw[cap=round,line width = 0.8mm,brown]  (7.5,1.0) -- (7.5,4.0) ;
\draw node[scale=1.3] at (7.8,2.8) { $B$};

\draw[cap=round,line width = 0.8mm,brown] (1.0,7.5) -- (4.0,7.5);
\draw[cap=round,line width = 0.8mm,brown]  (2.5,6.0) -- (2.5,9.0) ; 
\draw node[scale=1.3] at (7.8,7.8) { $B$};

\draw[cap=round,line width = 0.8mm,brown] (6.0,7.5) -- (9.0,7.5);
\draw[cap=round,line width = 0.8mm,brown]  (7.5,6.0) -- (7.5,9.0) ; 
\draw node[scale=1.3] at (2.8,7.8) { $B$};


\draw[green,cap=round,line width = 0.8mm] (2.5,4.2) -- (2.5,5.7);
\draw[red,cap=round,line width = 0.5mm]  (1.5,4.9) -- (3.5 ,4.9) ; 
\draw[red,cap=round,line width = 0.5mm]  (1.5,5.1) -- (3.5 ,5.1) ; 
\draw node[scale=1.3] at (2.8,5.4) { $A$};

\draw[green,cap=round,line width = 0.8mm] (7.5,4.2) -- (7.5,5.7);
\draw[red,cap=round,line width = 0.5mm]  (6.5,4.9) -- (8.5 ,4.9) ; 
\draw[red,cap=round,line width = 0.5mm]  (6.5,5.1) -- (8.5 ,5.1) ; 
\draw node[scale=1.3] at (7.8,5.4) { $A$};

\draw[green,cap=round,line width = 0.8mm] (4.2,2.5) -- (5.7,2.5);
\draw[red,cap=round,line width = 0.5mm]  (4.9,1.5) -- (4.9 ,3.5) ; 
\draw[red,cap=round,line width = 0.5mm]  (5.1,1.5) -- (5.1 ,3.5) ; 
\draw node[scale=1.3] at (5.3,2.8) { $A$};

\draw[green,cap=round,line width = 0.8mm] (4.2,7.5) -- (5.7,7.5);
\draw[red,cap=round,line width = 0.5mm]  (4.9,6.5) -- (4.9 ,8.5) ; 
\draw[red,cap=round,line width = 0.5mm]  (5.1,6.5) -- (5.1 ,8.5) ; 
\draw node[scale=1.3] at (5.3,7.8) { $A$};

\draw[red,cap=round,line width = 0.5mm]  (5.1,5.1) -- (6.1 ,5.1) ; 
\draw[red,cap=round,line width = 0.5mm]  (5.1,5.1) -- (5.1 ,6.1) ;
\draw[red,cap=round,line width = 0.5mm]  (3.9,5.1) -- (4.9 ,5.1) ;
\draw[red,cap=round,line width = 0.5mm]  (4.9,5.1) -- (4.9 ,6.1) ;
\draw[red,cap=round,line width = 0.5mm]  (5.1,3.9) -- (5.1 ,4.9) ;
\draw[red,cap=round,line width = 0.5mm]  (5.1,4.9) -- (6.1 ,4.9) ;
\draw[red,cap=round,line width = 0.5mm]  (3.9,4.9) -- (4.9 ,4.9) ;
\draw[red,cap=round,line width = 0.5mm]  (4.9,3.9) -- (4.9,4.9) ;
\draw node[scale=1.3] at (5.4,5.4) { $T$};

\draw[blue,cap=round,line width=0.5mm] (2.6,5.0) -- (5.0,5.0) -- (5.0,7.4) -- (2.6,7.4) -- (2.6,5.0);
\end{tikzpicture}
\caption{Construction of tensor $\mathcal{M}$ shown as the four tensors sharing the blue loop.  This is a possible single tensor which can be contracted with itself recursively to generate the partition function.}
\label{fig:trg}
\end{figure}
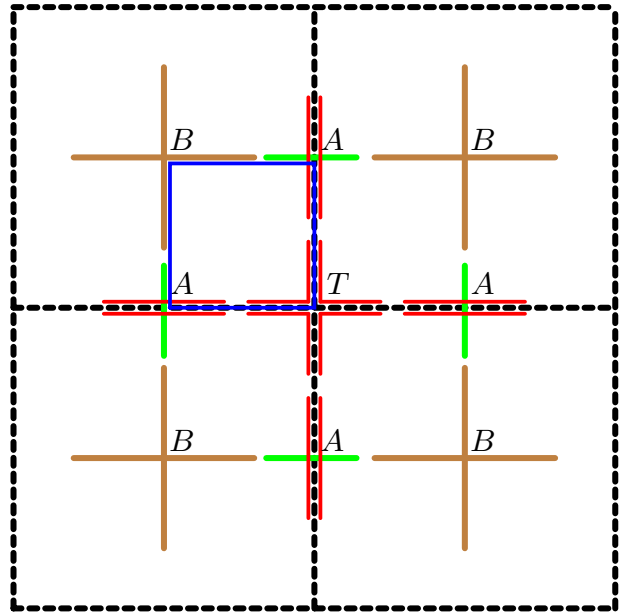

For this goal, we define a tensor, $ \mathcal{M} $, using a single elementary plaquette tensor (the $ B $), two link tensors  (the $A$s), and a single fermion $ T $ tensor.  This is shown diagrammatically in Fig.~\ref{fig:trg}.  As can be seen from the figure, there are two different types of indices associated with each direction in the tensor.  Each direction has one $m$ index, and one $K$ index.  However, repeated contraction of this tensor with itself in the appropriate pattern reproduces the partition function.  This is the only fundamental tensor necessary to do that.  The tensor is then explicitly given as,
\begin{align}
\label{eq:fund}
\nonumber
& \mathcal{M}_{m_1 m_2  m_3 m_4 K_{1} K_2 K_3 K_4 } =  
\sum_{m'_{1}, m'_{2}, \bar{K}_{1}, \bar{K}_{2}} B_{m_1 m^{'}_1 m_2 m^{'}_2} \times \\
& A_{m^{'}_2 m_3 K_{1} \bar{K}_1 }  T_{\bar{K}_1 K_2 \bar{K}_2 K_3} A_{m_4 m^{'}_1 \bar{K}_{2} K_{4}}.
\end{align}
Here the $K$ indices always have dimension four, however the $m$ indices run over all integers.  The $m$ indices are constrained by the $K$ indices though.  Looking at a single direction, the total size of the state-space associated with two of the indices is $D_{\text{bond}} = N_{\text{gauge}} \times 4$, where $N_{\text{gauge}}$ is the number of states allowed for the $B$ tensor index in practice.  

\subsection{Symmetric tensor}
It's possible to form a completely symmetric tensor in both space and time, as opposed to the asymmetric tensor constructed above.  This tensor formulation relies on ``dressing'' the link fermion tensors in their surrounding gauge field configurations.  This is possible because of how the $B$ tensor is completely diagonal in its four indices.

\begin{figure}
\centering
\begin{tikzpicture}[scale=0.5]
\draw node[scale=1.3] at (-9.6, -3) { $ m_1$};
\draw node[scale=1.3] at (-9.6, 3) { $ m_2 $};
\draw node[scale=1.3] at (-12, 0.3) { $m_3$};
\draw node[scale=1.3] at (-6, 0.3) { $m_4$};
\draw node[scale=1.3] at (-0.6, -3) { $ m_1$};
\draw node[scale=1.3] at (-0.6, 3) { $ m_2 $};
\draw node[scale=1.3] at (-3, 0.3) { $m_3$};
\draw node[scale=1.3] at (3, 0.3) { $m_4$};
\draw[cap=round,line width = 0.8mm,brown] (-9.0,-3.0) -- (-9.0,3.0);
\draw[cap=round,line width = 0.8mm,brown]  (-12.0,0.0) -- (-6.0,0.0) ;
\draw[cap=round,line width = 0.8mm,brown] (0.0,-3.0) -- (0.0, -2.0);
\draw[cap=round,line width = 0.8mm,brown] (0.0, 3.0) -- (0.0, 2.0);
\draw[cap=round,line width = 0.8mm,brown]  (-3.0,0.0) -- (-2.0,0.0) ;
\draw[cap=round,line width = 0.8mm,brown]  (3.0,0.0) -- (2.0,0.0) ;
\draw[cap=round,line width = 0.8mm,brown] (-2., -1.2) -- (-2., 1.2) ;
\draw[cap=round,line width = 0.8mm,brown] (2., -1.2) -- (2., 1.2) ;
\draw[cap=round,line width = 0.8mm,brown] (-1.2, -2.) -- (1.2, -2.) ;
\draw[cap=round,line width = 0.8mm,brown] (-1.2, 2.) -- (1.2, 2.) ;
\draw[cap=round,line width = 0.8mm,brown] (-1.4, 2.) -- (-2., 2.) -- (-2., 1.4) ;
\draw[cap=round,line width = 0.8mm,brown] (1.4, 2.) -- (2., 2.) -- (2., 1.4) ;
\draw[cap=round,line width = 0.8mm,brown] (-1.4, -2.) -- (-2., -2.) -- (-2., -1.4) ;
\draw[cap=round,line width = 0.8mm,brown] (1.4, -2.) -- (2., -2.) -- (2., -1.4) ;
\draw[->,black,ultra thick]  (-5.25, 0.0) -- (-3.75, 0.0) ;
\end{tikzpicture}
\caption{A graphical representation of how the decomposition of the $B$ tensor takes place.  Each smaller tensor is also diagonal so that all $m$ indices must take on the same values.}
\label{fig:Bsplit}
\end{figure}
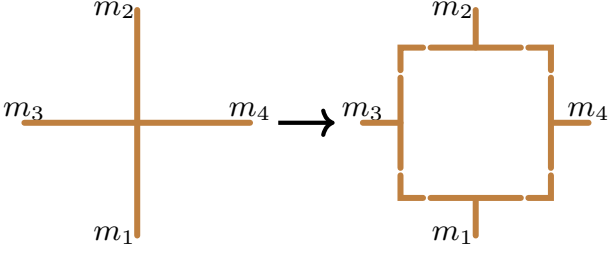

To construct the symmetric tensor, the first step is to separate the $B$ tensor into eight smaller pieces, four of which are associated with the adjacent link tensors, and the other four are associated with the four adjacent site tensors,
\begin{align}
\nonumber
  & B_{m_1 m_2 m_3 m_4} = \sum_{\alpha,\beta,\gamma,\sigma} b_{m_1 \sigma \alpha} b_{m_2 \alpha \beta} b_{m_3 \beta \gamma} b_{m_4 \gamma \sigma} \\
  &= \sum_{\alpha,\beta,\gamma,\sigma,\rho,\lambda,\chi,\psi}
  b_{m_1 \psi \alpha} \delta_{\alpha \beta} b_{m_2 \beta \gamma} \delta_{\gamma \sigma} b_{m_3 \sigma \rho} \delta_{\rho \lambda} b_{m_4 \lambda \chi} \delta_{\chi \psi}.
\end{align}
The $b$ tensors are also diagonal, and the $\delta$ matrices are simply Kronecker deltas.  This decomposition can be seen graphically in Fig.~\ref{fig:Bsplit}.  In principle, each of the above sums runs over all the integers; however, in practice one is forced to restrict the sum.

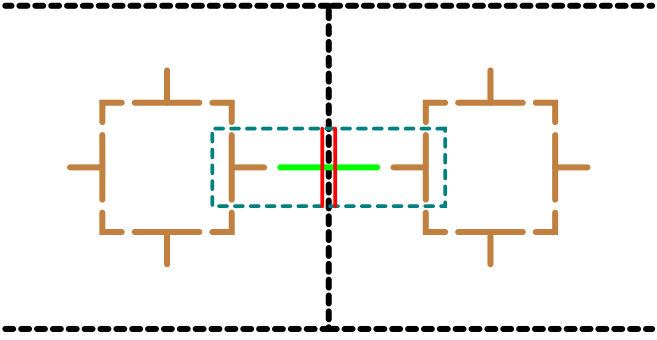
\begin{figure}
\centering
\begin{tikzpicture}[scale=0.86]
\draw[black, cap=round,line width=0.8mm, dashed] (0,0) -- (5,0) -- (5,5)  -- (10,5);
\draw[black, cap=round,line width=0.8mm, dashed] (5,0) -- (10,0);
\draw[black,cap=round,line width=0.8mm,dashed] (5,5) -- (0,5); 
\draw[cap=round,line width=0.8mm,brown] (1.0, 2.5) -- (1.5, 2.5);
\draw[cap=round,line width=0.8mm,brown] (4.0, 2.5) -- (3.5, 2.5);
\draw[cap=round,line width = 0.8mm,brown]  (2.5, 1.0) -- (2.5, 1.5) ;
\draw[cap=round,line width = 0.8mm,brown]  (2.5, 4.0) -- (2.5, 3.5) ;

\draw[cap=round,line width=0.8mm,brown] (6.0, 2.5) -- (6.5, 2.5);
\draw[cap=round,line width=0.8mm,brown] (9.0, 2.5) -- (8.5, 2.5);
\draw[cap=round,line width = 0.8mm,brown]  (7.5, 1.0) -- (7.5, 1.5) ;
\draw[cap=round,line width = 0.8mm,brown]  (7.5, 4.0) -- (7.5, 3.5) ;
\draw[cap=round,line width = 0.8mm,brown] (1.5, 2) -- (1.5, 3.) ;
\draw[cap=round,line width = 0.8mm,brown] (1.5, 3.2) -- (1.5, 3.5) -- (1.8, 3.5) ;
\draw[cap=round,line width = 0.8mm,brown] (2, 3.5) -- (3, 3.5) ;
\draw[cap=round,line width = 0.8mm,brown] (3.2, 3.5) -- (3.5, 3.5) -- (3.5, 3.2) ;
\draw[cap=round,line width = 0.8mm,brown] (3.5, 3.) -- (3.5, 2.) ;
\draw[cap=round,line width = 0.8mm,brown] (3.5, 1.8) -- (3.5, 1.5) -- (3.2, 1.5) ;
\draw[cap=round,line width = 0.8mm,brown] (3., 1.5) -- (2, 1.5) ;
\draw[cap=round,line width = 0.8mm,brown] (1.8, 1.5) -- (1.5, 1.5) -- (1.5, 1.8) ;

\draw[cap=round,line width = 0.8mm,brown] (6.5, 2) -- (6.5, 3.) ;
\draw[cap=round,line width = 0.8mm,brown] (6.5, 3.2) -- (6.5, 3.5) -- (6.8, 3.5) ;
\draw[cap=round,line width = 0.8mm,brown] (7, 3.5) -- (8, 3.5) ;
\draw[cap=round,line width = 0.8mm,brown] (8.2, 3.5) -- (8.5, 3.5) -- (8.5, 3.2) ;
\draw[cap=round,line width = 0.8mm,brown] (8.5, 3.) -- (8.5, 2.) ;
\draw[cap=round,line width = 0.8mm,brown] (8.5, 1.8) -- (8.5, 1.5) -- (8.2, 1.5) ;
\draw[cap=round,line width = 0.8mm,brown] (8., 1.5) -- (7, 1.5) ;
\draw[cap=round,line width = 0.8mm,brown] (6.8, 1.5) -- (6.5, 1.5) -- (6.5, 1.8) ;

\draw[teal, cap=round,line width=0.5mm, dashed] (6.8, 3.1) -- (3.2, 3.1) -- (3.2, 1.9) -- (6.8, 1.9) -- (6.8, 3.1);

 \draw[green,cap=round,line width = 0.8mm] (4.25,2.5) -- (5.75,2.5);
\draw[red,cap=round,line width = 0.5mm]  (4.9,1.9) -- (4.9 ,3.1) ; 
\draw[red,cap=round,line width = 0.5mm]  (5.1,1.9) -- (5.1 ,3.1) ; 
 \end{tikzpicture}
 \caption{The modified $\tilde{A}$ tensor (boxed in teal), built from the original $A$ tensor, and the $b$ tensors from the decomposition of the two $B$ tensors on the adjacent plaquettes.}
\label{fig:Atilde}
\end{figure}

The $b$ tensors are contracted with adjacent $A$ tensors, and the Kronecker deltas are moved to the surrounding site tensors.  The new $A$ tensors, $\tilde{A}$, are given by,
\begin{equation}
  \tilde{A}_{(m_1 K  m_2) ({m'}_1  K'  {m'}_2)} = \sum_{\alpha, \beta}
  b_{\alpha m_1 {m'}_1} A_{\alpha \beta K K'} b_{\beta m_2 {m'}_2}.
\end{equation}
This $\tilde{A}$ matrix is diagonal, since it is diagonal in all three sets of indices (the $K$s, and the $m$s) due to the aforementioned diagonal nature of the $B$ tensor and the already diagonal nature of the $K$ indices in the $A$ tensor.  This tensor can be seen in Fig.~\ref{fig:Atilde}.

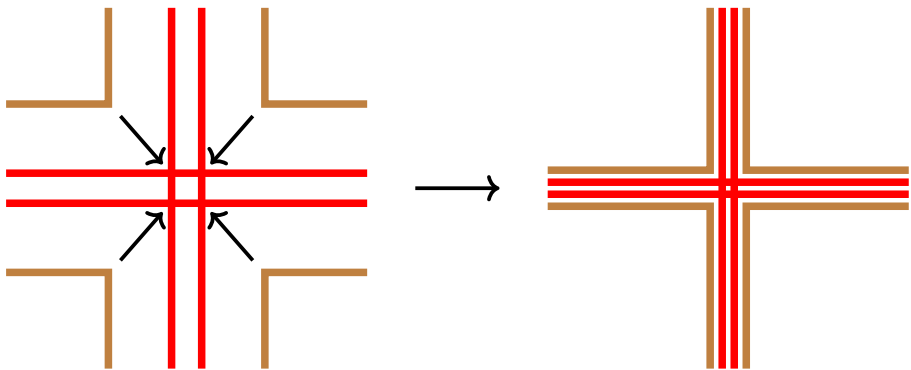
\begin{figure*}
\centering
\begin{tikzpicture}[scale=0.8]
\draw[red, line width=1mm]  (-9, 0.25) -- (-3, 0.25);
\draw[red, line width=1mm]  (-9, -0.25) -- (-3, -0.25);
\draw[red, line width=1mm]  (-6.25, 3) -- (-6.25, -3);
\draw[red, line width=1mm]  (-5.75, 3) -- (-5.75, -3);

\draw[brown, line width=1mm] (-9, -1.4) -- (-7.3, -1.4) -- (-7.3, -3) ;
\draw[brown, line width=1mm] (-9, 1.4) -- (-7.3, 1.4) -- (-7.3, 3) ;
\draw[brown, line width=1mm] (-3, 1.4) -- (-4.7, 1.4) -- (-4.7, 3) ;
\draw[brown, line width=1mm] (-3, -1.4) -- (-4.7, -1.4) -- (-4.7, -3) ;

\draw[->, line width=0.5mm] (-7.1, -1.2) -- (-6.4, -0.4);
\draw[->, line width=0.5mm] (-7.1, 1.2) -- (-6.4, 0.4);
\draw[->, line width=0.5mm] (-4.9, -1.2) -- (-5.6, -0.4);
\draw[->, line width=0.5mm] (-4.9, 1.2) -- (-5.6, 0.4);

\draw[->, line width=0.5mm] (-2.2, 0) -- (-0.8, 0);

\draw[red, line width=1mm]  (0, 0.1) -- (6, 0.1);
\draw[red, line width=1mm]  (0, -0.1) -- (6, -0.1);
\draw[red, line width=1mm]  (3.1, 3) -- (3.1, -3);
\draw[red, line width=1mm]  (2.9, 3) -- (2.9, -3);

\draw[brown, line width=1mm] (0, -0.3) -- (2.7, -0.3) -- (2.7, -3) ;
\draw[brown, line width=1mm] (0, 0.3) -- (2.7, 0.3) -- (2.7, 3) ;
\draw[brown, line width=1mm] (6, 0.3) -- (3.3, 0.3) -- (3.3, 3) ;
\draw[brown, line width=1mm] (6, -0.3) -- (3.3, -0.3) -- (3.3, -3) ;

\end{tikzpicture}
\caption{The modified fermion tensor.  The corners of the decomposed $B$ tensor are moved to the $T$ tensor at each site.  These corners are Kronecker deltas, and enforce that each site around a plaquette has the same plaquette quantum $m$ number.}
\label{fig:Ttilde}
\end{figure*}

For the site tensor ($T$ tensor), we now ``wrap'' it in Kronecker deltas which enforce that all four site tensors around a plaquette have the same $m$-plaquette number associated with that plaquette.  The new $\tilde{T}$ tensor has the form,
\begin{align}
\nonumber
  & \tilde{T}_{(m_1 K_1 m_8)(m_4 K_2 m_5)(m_2 K_3 m_3)(m_6 K_4 m_7)} = \\
  & T_{K_1 K_2 K_3 K_4} \delta_{m_1 m_2} \delta_{m_3 m_4} \delta_{m_5 m_6} \delta_{m_7 m_8},
\end{align}
and can be seen in Fig.~\ref{fig:Ttilde}.
At this point, there are no $B$ tensors remaining.  The partition function is simply a contraction of the $\tilde{A}$ and $\tilde{T}$ tensors.  To construct a single, symmetric, translation invariant tensor, we split the diagonal $\tilde{A}$ into two halves using the singular value decomposition,
\begin{align}
\nonumber
  \tilde{A}_{I J} & = \sum_{\alpha, \beta} U_{I \alpha} \lambda_{\alpha \beta} U^{\dagger}_{\beta J} \\ \nonumber
  & = \sum_{\alpha, \beta, \gamma} (U_{I \alpha} \sqrt{\lambda_{\alpha \beta}})(\sqrt{\lambda_{\beta\gamma}} U^{\dagger}_{\gamma J}) \\
  & = \sum_{\alpha} L_{I \alpha} {L^{\dagger}}_{\alpha J}.
\end{align}
Furthermore, there are singular values with value zero, and they can be removed to decrease the size of the state space.  This is equivalent to taking the square-root of the $\tilde{A}$ matrix and removing the zero columns (rows).  With the $L$ matrices we can now form a symmetric tensor, by contracting four of these matrices with a $\tilde{T}$,
\begin{equation}
  S_{ijkl}(\beta) = \sum_{\alpha,\beta,\gamma,\delta} 
  \tilde{T}_{\alpha \beta \gamma \delta} L_{\alpha i} L_{\beta j} L_{\gamma k} L_{\delta l}.
\end{equation}
This tensor is symmetric in space and time, and since the $L$ matrices are diagonal, its nonzero tensor elements are constrained by the fermion tensor, $T$. This final $S$ tensor satisfies the same constraint as the original fermion $T$ tensor, however with tensor elements with values other than $1$, instead given by linear combinations of modified Bessel functions which are functions of the gauge coupling.

\section{Numerical Simulation: HOTRG and HMC}

We implemented the HOTRG algorithm to evaluate $\ln Z$ using the tensor defined in Eq.~ \eqref{eq:fund} as a translation invariant tensor for coarse-graining. We measured the average plaquette,
\begin{equation}
    \langle U_{p} \rangle = \frac{1}{N_{s}N_{\tau}} \frac{\partial \ln Z}{\partial \beta},
\end{equation}
as a function of the gauge coupling and compared it to numerical data from Ref.~\cite{GOSCHL201763}. In this case our computation using HOTRG completely agrees with the worm algorithm generated data. Moreover we can add a $ \theta $ term to the original action which results in new couplings, expressed as linear combinations of the gauge coupling and theta parameter, $ \eta = \frac{\beta}{2} - \frac{\theta}{4 \pi} $ and $ \bar{\eta} =  \frac{\beta}{2} + \frac{\theta}{4 \pi} $. For the tensor construction here we only need to redefine the plaquette tensor, $B$, with $ I_{m} (\beta) $ replaced by $ I_{m} (2 \sqrt{\eta\bar{\eta}}) \left({\eta / \bar{\eta}} \right)^{m/2} $.

\begin{figure}[tbp]
\centering
\includegraphics[width=8.6cm]{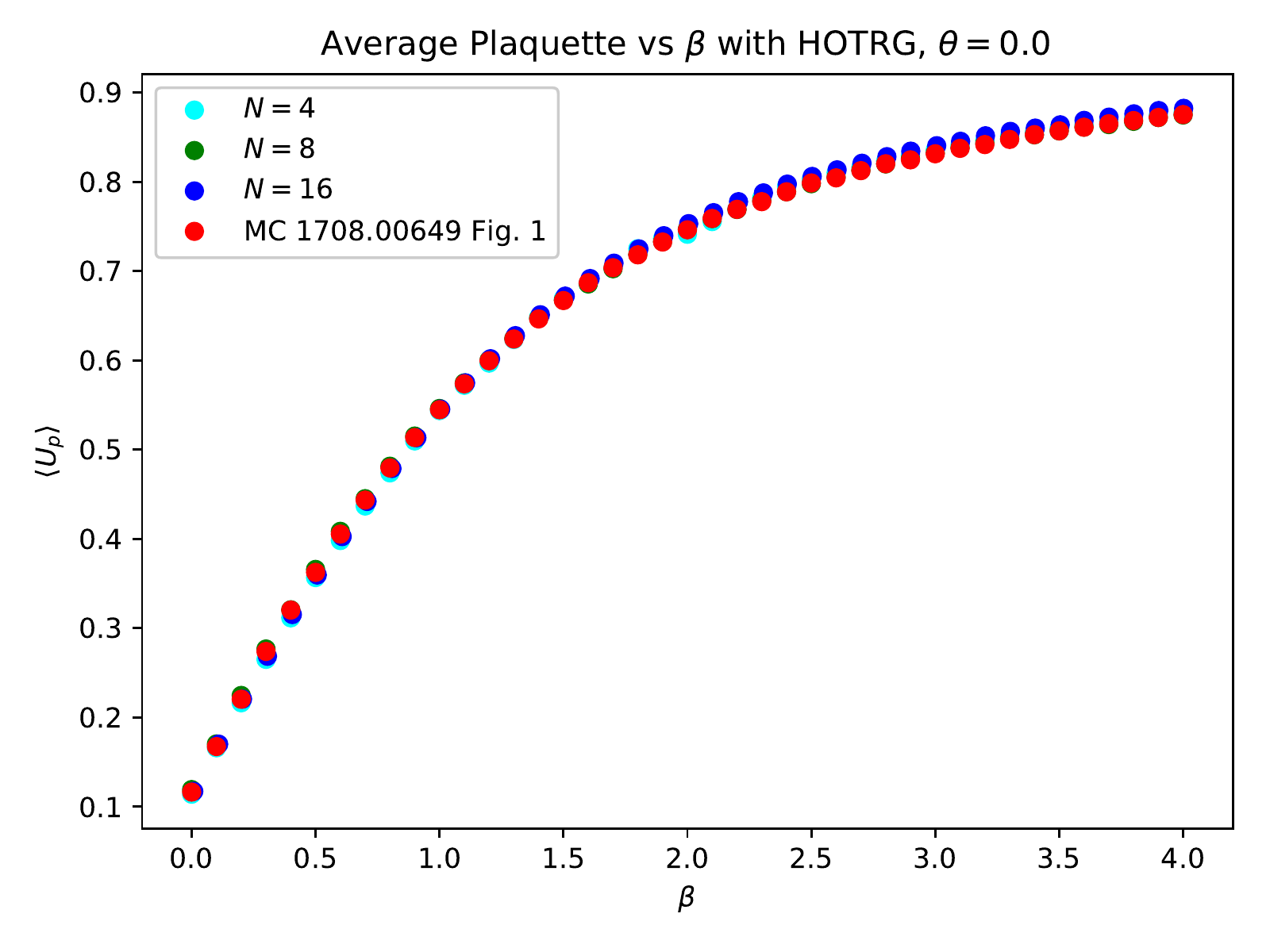}
\caption{Average Plaquette vs. $ \beta $ for lattice sizes with $N_{\tau} = N_{s} = 4$, $8$, and $16$ and compared with data from Ref.~\cite{GOSCHL201763}.  For this data $N_{\text{gauge}} = 3$ is sufficient to achieve similar accuracy to MC data.}
\label{fig:avg_plaq}
\end{figure}

\begin{figure}[tbp]
\centering
\includegraphics[width=8.6cm]{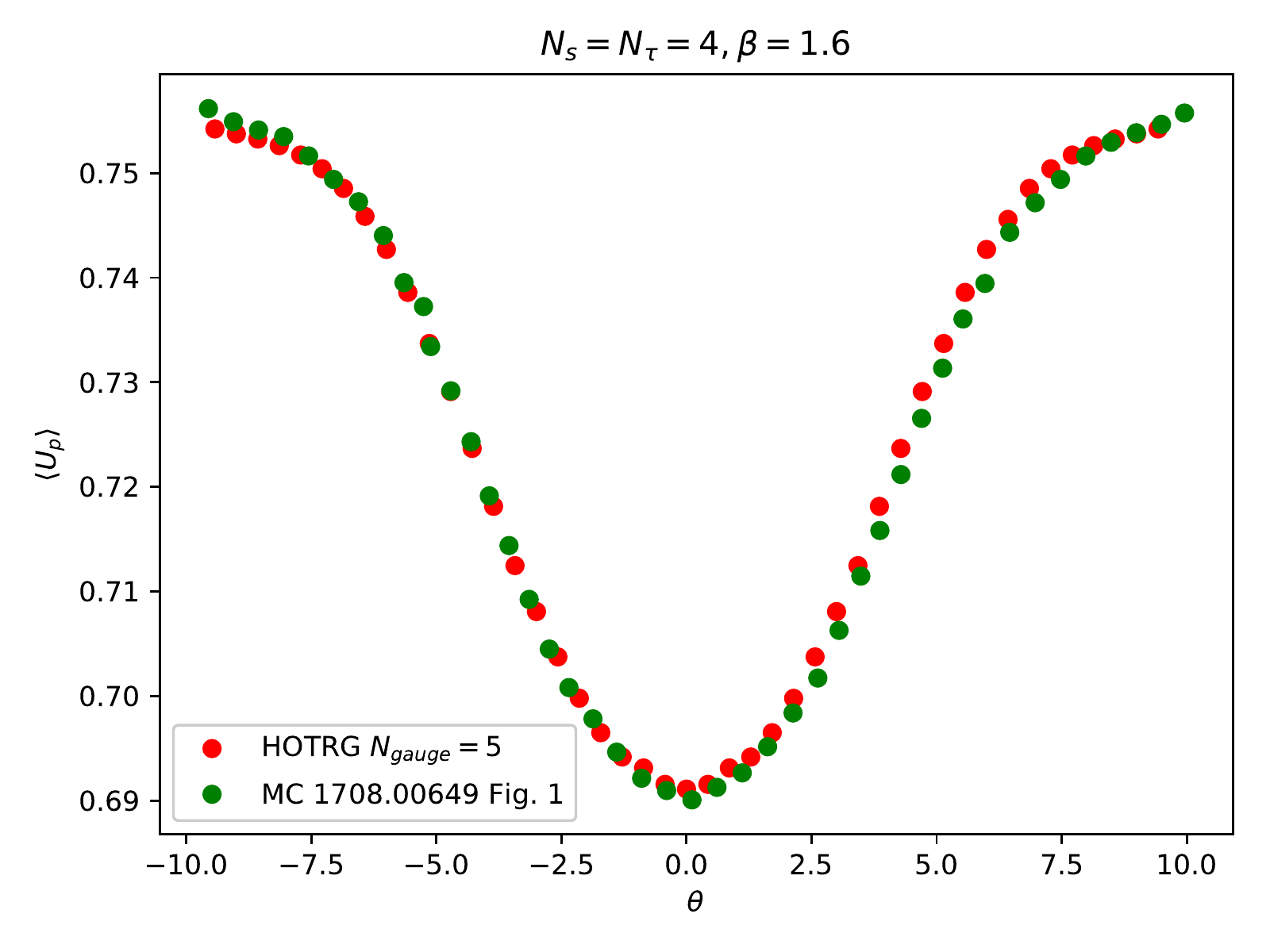}
\caption{Average Plaquette vs. $ \theta $ for a lattice with $N_{s} = N_{\tau} = 4$.  Here $N_{\text{gauge}} = 5$ is necessary to achieve similar accuracy to the MC data.}
\label{fig:plaqvstheta}
\end{figure}

\begin{figure}
    \centering
    \includegraphics[width=8.6cm]{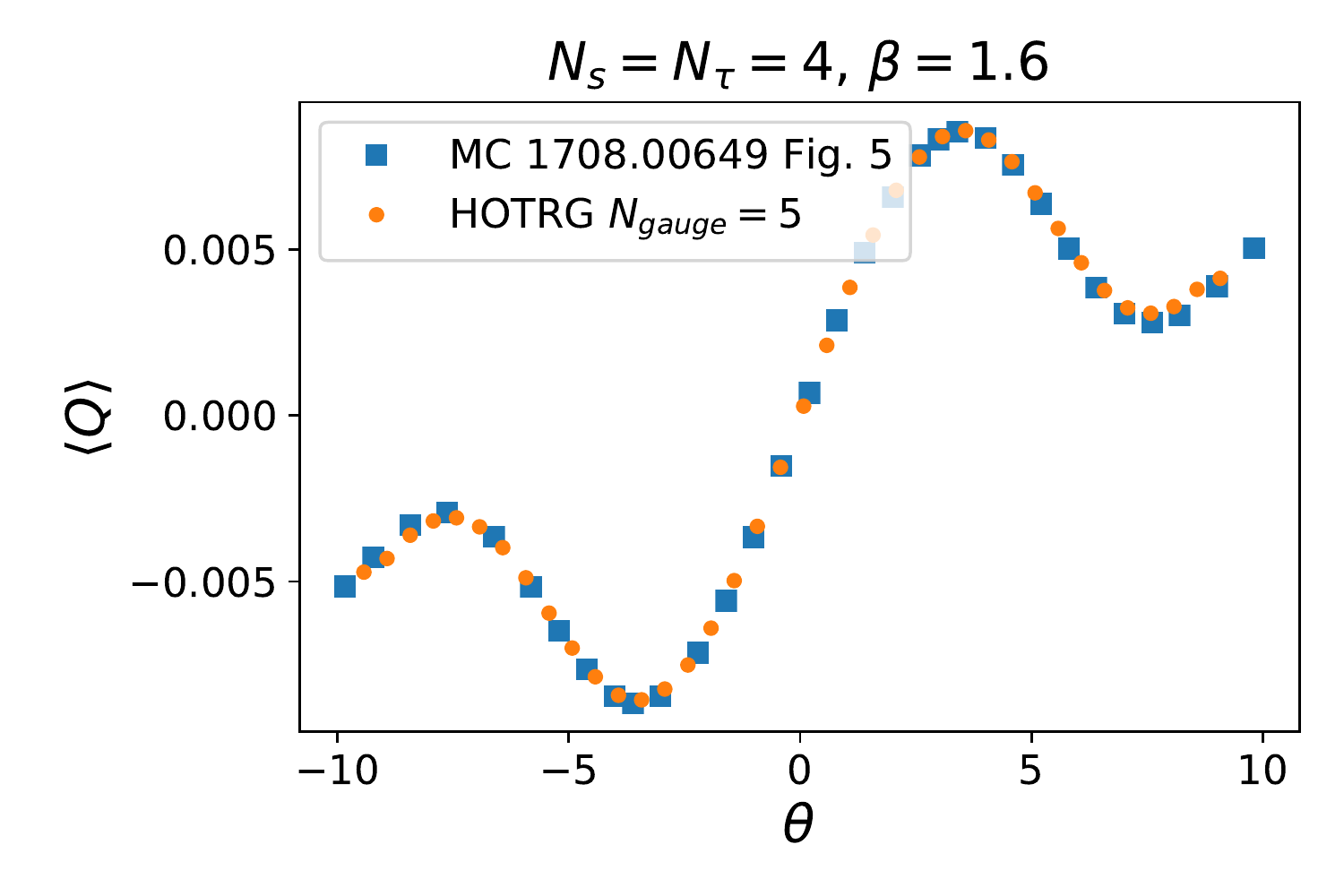}
    \caption{The topological charge as a function of $\theta$.  Here we compare with Ref.~\cite{GOSCHL201763}.  We find a slightly larger range of plaquette quantum numbers are necessary---in contrast to the average plaquette---to achieve consistent results.  In this case, the plaquette numbers had to be allowed to run from $m=-2$ to $2$.}
    \label{fig:topo_charge}
\end{figure}

To ensure the formulation is valid, we measured a couple of observables, including the average plaquette $ \langle U_{p} \rangle $, and the topological charge, $\langle Q \rangle$ as a function of the $\theta$ parameter.  The topological charge is defined as,
\begin{equation}
    \langle Q \rangle = \frac{1}{N_{s}N_{\tau}} \frac{\partial \ln Z}{\partial \theta}.
\end{equation}
The results of the calculation of the average plaquette as a function of $\beta$ for different system sizes can be seen in Fig.~\ref{fig:avg_plaq},
and as a function of $\theta$ in Fig.~\ref{fig:plaqvstheta}.
We find good agreement and convergence across a wide range of $\beta$ values for the relatively small number of gauge states, $N_{\text{gauge}} = 3$ and 5.  The results for the topological charge can be seen in Fig.~\ref{fig:topo_charge},
and again we find good agreement across the range scanned, however to obtain this result a larger $N_{\text{gauge}} = 5$ was necessary.  We also noted that for large $\theta$ values, data on larger volumes was significantly more noisy.  We discuss possible explanations and solutions in the conclusions.

\section{Conclusions}
In this paper we have constructed a tensor network formulation of the massless
lattice Schwinger model with staggered fermions. We have
considered both the usual action and one in which a topological
term is added. The addition of the latter term 
induces a sign problem and renders the
model intractable for a conventional hybrid Monte Carlo simulation.

Using the HOTRG algorithm we
have computed the free energy and its derivatives and compared the results,
where possible,
with both hybrid Monte Carlo simulations and simulations based on
a dual representation based on fermion loops. Where comparison is possible
the agreement is good with the tensor network calculation being
superior computationally to Monte Carlo.  That said, we have experienced
difficulties measuring observables for large values of the topological coupling
$\theta$.  Typically the signal for an operator like the plaquette becomes
very noisy after several iterations of the blocking scheme. 

Additionally, arguments used for the positivity of terms in the sum of the partition function assume a complete lattice with boundary conditions and lattice size already achieved~\cite{GATTRINGER2015732}.  In contrast, the HOTRG does not know before-hand what the final size of the lattice will be, or what the boundary conditions will be at that size.  This in turn gives the algorithm more freedom to choose which states are relevant during truncation, even though those very states may be projected out in the final step of blocking; making them useless.

A tensor construction scheme which uses an environment tensor might achieve better results at larger volumes, since, the forward-backward iteration from a complete lattice should retroactively adjust the intermediate states kept during truncation at smaller volumes.

Of course in the continuum limit the partition function should be
independent of $\theta$  and the difficulties are likely related at least in
part to this fact --- as the chiral symmetry of the lattice action is restored
the system will develop chiral zero modes which will suppress the contribution
of any topological field configurations to the partition function.

The $\theta$ dependence is restored in the presence of a fermion mass.
However in that case there are non-trivial $-1$ factors which appear in the dual
representation of the partition function. Part of the phase depends on the
number of closed fermion loops appearing in any particular dual configuration. It is extremely hard to see how this phase can be reconstructed from the contraction of local tensors and we have not been able to generalize the tensor
network described here to the case of non-zero masses. This should sound a cautionary note to the idea that tensor network formulations of lattice
field theories are free of sign problems. In the case of fermion
theories this may not be generically the case.

\begin{acknowledgments}The authors would like to thank the members of the QuLat collaboration for stimulating discussions.  SC, YM, and JUY were supported by  the  U.S.  Department  of  Energy  (DOE)  under  Award  Number DE-SC0019139. \end{acknowledgments}


\end{document}